\documentclass[prl,twocolumn,aps,showpacs,preprintnumbers,amsmath,amssymb]{revtex4}
\usepackage{graphicx}
\usepackage{dcolumn}
\usepackage{bm}
\usepackage{amsmath} 

\begin{document} 
\title{Dynamics of $F=1$ $^{87}$Rb  condensates at finite temperatures}

\author{J. Mur-Petit$^1$, M. Guilleumas$^1$, A. Polls$^1$,
 A. Sanpera$^{2,}$, and M. Lewenstein$^{3,4}$}

\affiliation{$^1$Departament d'Estructura i Constituents de la Mat\`eria,
Facultat de F\'isica,\\ 
Universitat de Barcelona, E-08028 Barcelona, Spain\\
$^2$ ICREA and Grup de F\'isica Te\`orica, Universitat Aut\`onoma de 
Barcelona, E-08193 Bellaterra, Spain\\
$^3$ ICREA and ICFO-Institut de Ci\`encies Fot\`oniques, E-08034 Barcelona, Spain\\
$^4$ Institut f\"ur Theoretische Physik, Universit\"at Hannover, D-30167
Hannover, Germany}

\author{K. Bongs$^5$, and K. Sengstock$^5$}
\affiliation{
$^5$ Institut f\"ur Laser-Physik, Universit\"at Hamburg,
D-22761 Hamburg, Germany}

\date{\today}


\begin{abstract}
We investigate the dynamics of a $F=1$ spinor Bose-Einstein condensate of
$^{87}$Rb atoms confined in a quasi-one-dimensional trap both at zero 
and at finite temperature. 
At zero temperature, 
we observe coherent oscillations between populations of the various spin
components and formation of multiple domains in the condensate. 
We study also finite temperature effects in the spin dynamics 
taking into account the phase fluctuations in the Bogoliubov-de Gennes
framework. At finite $T$, despite complex 
multi-domain formation in  the condensate, 
population equipartition occurs. 
The length scale of these spin domains seems 
to be determined intrinsically  by non-linear interactions. 
\end{abstract}


\maketitle

\section{Introduction}

The seminal theory papers of T.-L. Ho \cite{Ho98} and T. Ohmi and K. Machida \cite{machida},  as well as the experiments 
performed by the MIT group on optically trapped sodium 
Bose-Einstein condensates (BEC) 
\cite{Sten98} have stimulated the development of a 
new interesting  area of research in the field of 
multi-component quantum gases: BEC with spin degrees of freedom, 
i.e. the so called {\it spinor condensates}.
The bosonic quantum field operator in such systems is no longer a scalar, 
but  a vector, and depending on various parameters 
the system can be found in a coherent superposition or in an  
incoherent mixture of condensates with different spin components, that can
exchange population depending on the intrinsic nonlinear coupling. 
The couplings between the different spin components, similarly as 
in the standard scalar BEC can be described by zero range potentials 
that however, acquire a matrix form and describe couplings that assure 
total spin conservation in elastic binary collisions \cite{review}.  

It is worth mentioning that spinor condensates are closely related
to the effective spin-1/2 systems realized 
by radio-frequency coupling of the two
hyperfine states in $^{87}$Rb \cite{spin1o2}; in such systems 
spin-waves above the critical temperature \cite{McGuirk} and
decoherence effects were observed \cite{Lewandowski}.
From the theoretical point of view, collective modes in
two-component BEC's have been studied in \cite{collective}, while
an analysis on the formation of spin domains can be found e.g. in
\cite{domain}.

Spinor condensates are suitable systems to study various quantum phenomena, 
that do not occur in single component BEC's. 
Equilibrium states of 
spinor condensates in an optical trap can exhibit magnetic ordering 
of various kinds. For instance, for sodium ($F=1$), 
the freedom of spin orientation leads to the formation of 
spin domains in an external 
magnetic field, which can be either miscible or immiscible 
with one another \cite{Sten98}.
Excitations in such systems may manifest the different spin character;
stability of topological 
excitations and textures, such as ordinary vortices, 
coreless vortices (Skyrmions) \cite{Ho98,stoof-skyrmion}, 
and t'Hooft-Polyakov monopoles \cite{stoof}, 
depends in a complex manner on the parameters of the system. The complexity of these systems 
becomes even more transparent 
in the limit of strongly correlated systems:  
ultra-cold Bose spin gases in optical lattices exhibit fascinating properties, including new
types of quantum phases, such as a polar condensate phase, a condensate of 
singlet pairs, a crystal spin nematic phase, and a spin singlet crystal phase \cite{zhenia}.

The ground state, its magnetic properties and 
the low temperature thermodynamics of spinor condensates have been studied 
in several experiments: in $^{23}$Na \cite{Na} ($F=1$)
which has an anti-ferromagnetic ground state,
in $^{87}$Rb in the $F=1$ spin state \cite{Schmaljohann04,Chang04,Barrett01,Kuwamoto04}
which is ferromagnetic, as well as in the $F=2$ spin state, 
which presents a rich ground state behavior 
\cite{Schmaljohann04,Kuwamoto04}. 
Recent experiments involving non-destructive imaging of magnetization
of a spin $F=1$ Bose gas of $^{87}$Rb atoms \cite{Higbie05} open a 
new route to study magnetism of spinor condensates.
Also, the recent experiment with 
chromium \cite{tilman} condensates opens the possibility 
to study even higher spin states ($F=3$) with 
long range (dipolar)  interactions 
and, presumably a much more complex phase diagram.

In the last two years the focus of investigation has shifted towards 
two other very important aspects of the spinor BEC physics: the equilibrium
properties at finite $T$, and dynamics of spinor BEC's. 
The thermodynamic properties of an ultra-cold  spin Bose gas have been
experimentally investigated in Ref. \cite{Tfinita}.
Recently, finite temperature effects 
to describe the properties of the equilibrium density distribution 
have been considered within the Hartree-Fock-Popov theory \cite{Zhang04}.
Spin dynamics has been experimentally studied in $^{87}$Rb 
condensates for $F=1$ in \cite{Chang04}, and more exhaustively 
for $F=2$ in \cite{Schmaljohann04,Chang04,Kuwamoto04}. Coherent collisional spin dynamics in $^{87}$Rb for $F=2$
has been recently observed also 
in optical lattices \cite{lastbloch} - these experiments pave way towards
the efficient 
creation of entangled atom pairs in optical lattices.

Here we focus on the dynamics of  $F=1$ elongated 
spinor condensates at zero and finite temperatures.
One of our major motivation to study the thermal effects on spin dynamics
is to learn something about decoherence in multi-component systems.
In the case $F=1$ the internal coupling of the spin 
components  depends only on 2 coefficients, while for $F=2$ it depends on 3.
Due to this simplicity, the case $F=1$ allows for a better 
understanding of the interplay between nonlinear interactions, 
spin couplings and temperature effects. 
We consider here the full coupled dynamical equations of the spin
components, obtained within a mean-field framework, without
any further approximation such as the Single Mode Approximation (SMA)
\cite{Pu99,Yi02}, or variational ansatz \cite{Zhang05},
that would mask some aspects of the complex dynamical evolution. 
We restrict our analysis to the quasi-1D case, 
when the condensates are kinematically frozen in the transverse directions. 
We investigate then the influence of thermal effects on the spin dynamics. 
To this aim we use the Bogoliubov-de Gennes description 
of phonon modes and treat them as 
classical random fields, similarly as in Ref. \cite{Dettmer01}.
For quasi-1D condensates at finite $T$ the main 
contribution to phonon fluctuations comes from the fluctuations of the condensate phase \cite{Petrov00}, and 
this contribution is 
fully accounted in our simulations. 

The paper is organized as follows. First, in section II 
we introduce the model for $T=0$. We describe
some of the details of the numerical method in section III. 
In section IV we describe 
our results for $T=0$. 
Section V we present a discussion of finite temperature effects. 
%
We conclude in section VI.

\section{Description of the system}

The many-body Hamiltonian describing a F=1 spinor condensate 
in absence of an external magnetic field is 
given by \cite{Ho98}
\begin{eqnarray}
  H& = &\int d^3r \left\{
    \Psi^\dagger_m \left(-\frac{\hbar^2}{2M}{\bm \nabla}^2
                   + V_{ext}\right) \Psi_m \right.   \\
    &&
\left.
    +\frac{c_0}{2} \Psi^\dagger_m \Psi^\dagger_{j} \Psi_{j} \Psi_m
    +\frac{c_2}{2}
      \Psi^\dagger_m\Psi^\dagger_{j} {\bf F}_{mk} \cdot 
                            {\bf F}_{jl} \Psi_{l}\Psi_k 
\right\} 
\nonumber
\label{hamiltonian}
\end{eqnarray}
where $\Psi_m({\bf r})$ $(\Psi_m^\dagger)$ is the field 
operator that annihilates (creates) an atom in the $m$-th 
hyperfine state $|F=1, m \rangle$ at point ${\bf r}$
($m=1,0,-1$).
The trapping potential $V_{ext}({\bf r})$ is assumed harmonic
and spin independent.
The terms with coefficients $c_0$ and $c_2$
describe binary elastic collisions of spin-1 atoms
in the combined symmetric channel of total spin 0 and 2, 
and are expressed in
terms of the s-wave scattering lengths $a_0$ and $a_2$:
$c_0=4 \pi \hbar^2 (a_0+2a_2)/3M$ and
$c_2=4 \pi \hbar^2 (a_2-a_0)/3M$, where
$M$ is the atomic mass. ${\bf F}$ are the spin-1 matrices 
\cite{Ho98,Zhang04}.
The system is ferromagnetic if $c_2 < 0$ ($^{87}$Rb), 
and anti-ferromagnetic if $c_2 > 0$ ($^{23}$Na).

The total number of atoms 
$N=\int d{\bf r} (|\Psi_1|^2+|\Psi_0|^2+|\Psi_{-1}|^2)$
and the total magnetization
${\cal M}=\int d{\bf r} (|\Psi_1|^2-|\Psi_{-1}|^2)$
commute with the Hamiltonian, 
and are thus constants of motion.

In the mean field approach, a condensate order parameter is
introduced for each magnetic sublevel 
$\psi_m({\bf r})=\langle \Psi_m({\bf r}) \rangle$,
and by neglecting quantum fluctuations it yields the following
energy functional
\begin{eqnarray}
  {\cal E} &=& \int d^3r \left\{ 
    \psi^*_m \left(-\frac{\hbar^2}{2M}{\bm \nabla}^2
                   + V_{ext}\right) \psi_m \right.   \nonumber\\
    &&
    +\frac{c_0}{2} \psi^*_m \psi^*_{j} \psi_{j} \psi_m
    +\frac{c_2}{2} 
      \psi^*_m\psi^*_{j} {\bf F}_{mk} \cdot
                            {\bf F}_{jl} \psi_{l}\psi_k \,.
\label{functional}
\end{eqnarray}
According to $i \hbar \partial \psi_m/\partial t=
\delta {\cal E}/\delta \psi_m^*$, the 
coupled dynamical equations for the spin components are obtained
\cite{Pu99,Zhang03}
\begin{equation}
i \hbar \frac{\partial \psi_m}{\partial t} =
 \left[-\frac{\hbar^2}{2M}{\bm \nabla}^2 +V^{eff}_{m}\right]\psi_{m}
           + c_2 T_{m}^*  \,,
\label{dyneqs2}
\end{equation}
$n_m({\bf r})=|\psi_m|^2$ is the density of the $m$-th component and
$n=|\psi_{1}|^2+|\psi_0|^2+|\psi_{-1}|^2$ is
the total density normalized to the total number of atoms $N$.
The population of the hyperfine state $|1, m \rangle$
is $N_m=\int d{\bf r} |\psi_m|^2$ such that
$N=N_{1}+N_0+N_{-1}$.
We have defined $T^*_{\pm 1}=\psi_0^2 \psi_{\mp 1}^*$,
$T^*_{0}=2 \psi_{1} \psi_0^* \psi_{-1}$, and the
effective potentials that will determine the spatial dynamics
of each component
\begin{eqnarray}
V^{eff}_{\pm 1}&=&V_{ext}+ c_0 n+c_2(\pm n_{1}+n_0\mp n_{-1})
      \nonumber \\
V^{eff}_{0}&=&V_{ext}+ c_0 n+ c_2(n_{1}+n_{-1})  \,.
\label{veff}
\end{eqnarray}
Analogously to a spin-polarized condensate, the 
multicomponent Gross-Pitaevskii equations (\ref{dyneqs2})
can be rewritten in the form of continuity equations
\cite{Dalfovo99}:
\begin{equation}
\frac{\partial}{\partial t} n_m+ {\bm \nabla} \cdot {\bf j}_m=
\delta \dot{n}_m({\bf r},t)
 \,, \\
\label{continuity}
\end{equation}
where ${\bf j}_m=\hbar (\psi_m^* {\bm \nabla} \psi_m-
\psi_m {\bm \nabla} \psi_m^*)/2 i M$ is the current, and
$\delta \dot{n}_m({\bf r},t)=-(2 c_2/\hbar) {\rm Im}(T_m \psi_m)$
is the rate of transfer of populations between spin components.
Since the total number of atoms and magnetization are
conserved, and the dynamical
equations for $m=\pm 1$ are symmetric,
it is verified that
$\delta \dot{n}_0=-2\delta \dot{n}_{\pm 1}=
(2 c_2/i \hbar) \,(\psi_{1} \psi_0^{*2} \psi_{-1}-
\psi_{1}^* \psi_0^2 \psi_{-1}^*)$.

In our calculation,
we assume an axially-symmetric harmonic confinement 
$V_{ext}=M(\omega_{\perp}^2 r_{\perp}^2+\omega_z^2 z^2)/2$.
In the limit of highly elongated traps 
($\omega_{\perp} \gg \omega_z$) the tight confinement ensures
that no excited states are available in the transverse direction 
and thus the dynamics takes place along the axial direction.
Factorizing $\psi_m({\bf r})$ into a longitudinal and a transverse
function, and approximating the transverse part as the two-dimensional
ground state of the transverse oscillator \cite{Pu99,Zhang05}, the
equations of motion (\ref{dyneqs2}) become one-dimensional
for the longitudinal wave functions $\psi_m(z)$, and
the coupling constants $c_0$ and $c_2$ 
are accordingly rescaled by a factor $1/(2 \pi a_\perp^2)$,
with $a_\perp=\sqrt{\hbar/ m \omega_\perp}$ the
transverse oscillator length \cite{Olshanii98}.

\section{Numerical method}

The dynamics of the spin components is obtained by numerically
integrating the coupled nonlinear differential equations (\ref{dyneqs2}).
For the time evolution our numerical procedure 
combines the split operator method with the Fast Fourier Transform to treat
the kinetic terms and a fourth-order Runge-Kutta method
for the remaining terms of the dynamical equations.
We have compared our combined numerical method with an evolution using
a pure fourth-order Runge-Kutta as the one used in Ref.\cite{Pu99},
and we have found that our method
allows larger time steps making the computation notably faster.

In this paper, 
we consider $N=20000$ atoms of spin-1 $^{87}$Rb trapped in a 
quasi-1D harmonic trap with
$\omega_{\perp}=2\pi\times 891$Hz and $\omega_z=2\pi\times 21$ Hz.
The coupling constants are $a_0=101.8 a_B$, and $a_2=100.4 a_B$
\cite{vanKempen02}, with $a_B$ the Bohr radius. Since $c_2 <0$
the atomic interactions for $^{87}$Rb atoms are ferromagnetic.
The initial wave functions
$\psi_m(z,t=0)$ have the same spatial profile as the
ground state of the scalar condensate with coupling constant
$c_0/(2 \pi a_\perp^2)$. Each initial wave function component is however 
normalized to the corresponding initial population in spin-$m$, $N_m$.
Since $\psi_m(z,t)$ are complex functions, 
$\psi_m(z,0)=|\psi_m(z,0)| \exp{(i \theta_m)}$,
the initial phases have also to be fixed.
However, for initial symmetric configurations ($N_{1}=N_{-1}$)
the dynamical evolution depends only on one initial relative phase
$\theta=2\theta_0-\theta_1-\theta_{-1}$ \cite{Pu99},
and thus only one initial phase has to be fixed.
Experimentally the initial phases
are not well determined and the results are averaged 
over several repeated measurements \cite{Chang04}.
Therefore we will average the dynamics 
over different initial relative phases
randomly distributed over $(0,2\pi)$ in order to obtain results easy
to compare with experiments.

\section{Spinor dynamics at $T=0$}

\begin{figure}
\includegraphics[width=0.85\columnwidth,
angle=0, clip=true]{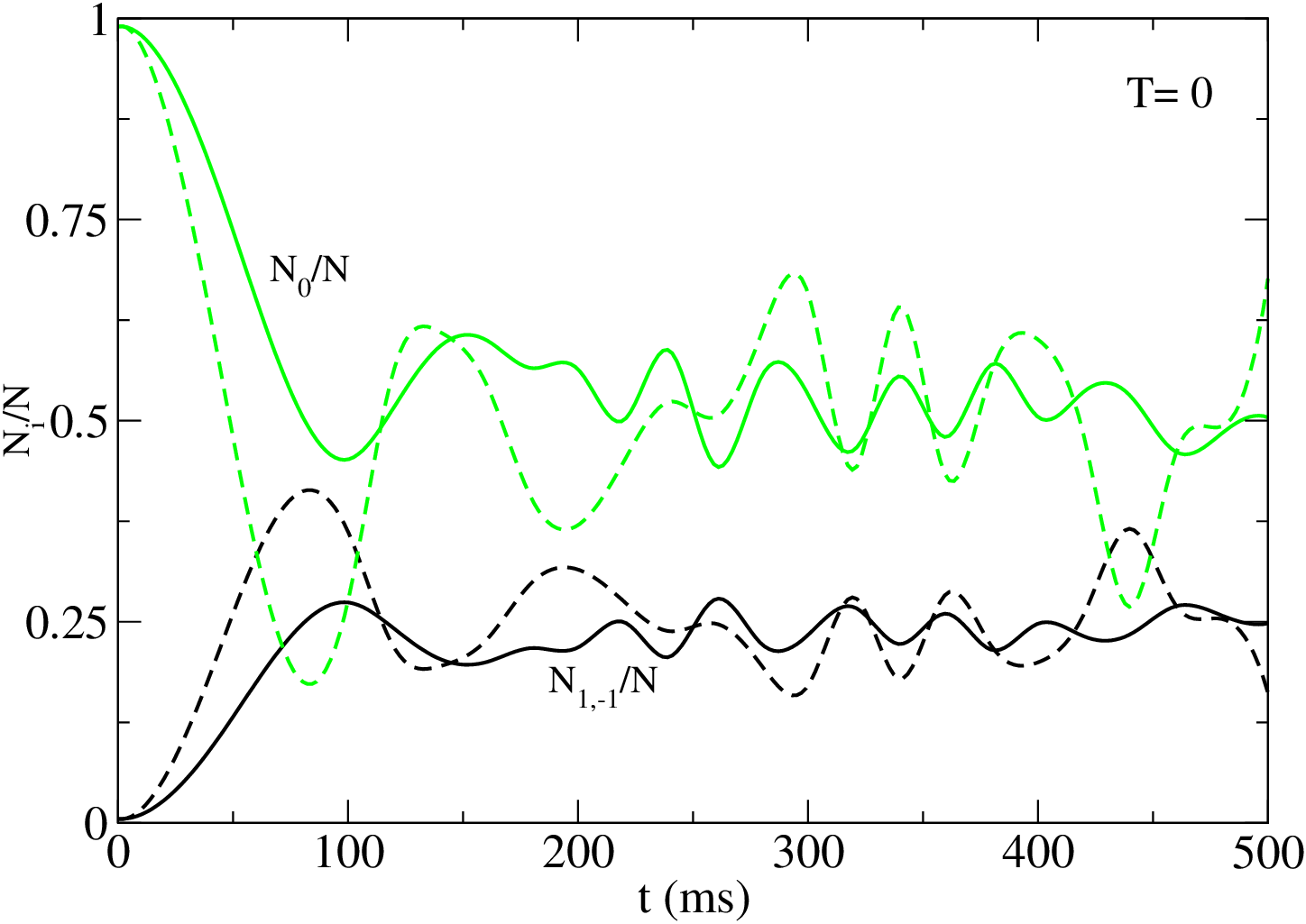}
\caption[]{(Color online) Population of the spin components as a function
of time for the initial configuration 
$(N_1/N,N_0/N,N_{-1}/N)=(0.5 \%,99 \%,0.5 \%)$ at $T=0$.
Solid lines: numerical results averaged over 20 random
initial relative phases $\theta$. Dashed lines: initial phase
$\theta=0$.
}
\label{fig1}
\end{figure}

We consider that initially a quasi-pure condensate in the
$m=0$ spin component is populated. 
Spin component mixing requires, at least, a small seed of atoms populating 
the other components but keeping the total magnetization equal to zero.  
In Figure \ref{fig1} we plot the population of each spin component
as a function of time at zero temperature, for the
initial populations $(N_1/N,N_0/N,N_{-1}/N)=(0.5 \%,99 \%,0.5 \%)$.
In absence of an external magnetic field gradient the
dynamical evolutions of the spin $m= \pm 1$ components are symmetric.
Dashed lines correspond to the dynamical evolution with a given
initial relative phase $\theta=0$,
and solid lines
are the average over $20$ random initial relative phases.
The oscillations of the populations with a given initial phase,
are smeared out by the average over different runs.
It is interesting to point out 
the damping of the oscillations
obtained in our numerical calculations
which is a finite size effect related 
to collapse phenomena characteristic of discrete anharmonic spectra
\cite{collapse}, and to dephasing of Josephson oscillations \cite{villain}. 
During the time evolution the magnetization is
conserved, as it has been experimentally observed \cite{Chang04}.
The populations oscillate around the ground state configuration 
of the system with ${\cal M}=0$ that
in absence of applied external magnetic field is
$(25\%,50\%,25\%)$ \cite{Zhang03}.
Our numerical results are in qualitative agreement with the
experimental measurements of Ref.\cite{Chang04} obtained in
a strongly anisotropic disk-shaped spin-1 condensate, where
the relaxation to the steady state is also not monotonic but damped.
 
\begin{figure}
\includegraphics[width=0.85\columnwidth,
angle=0, clip=true]{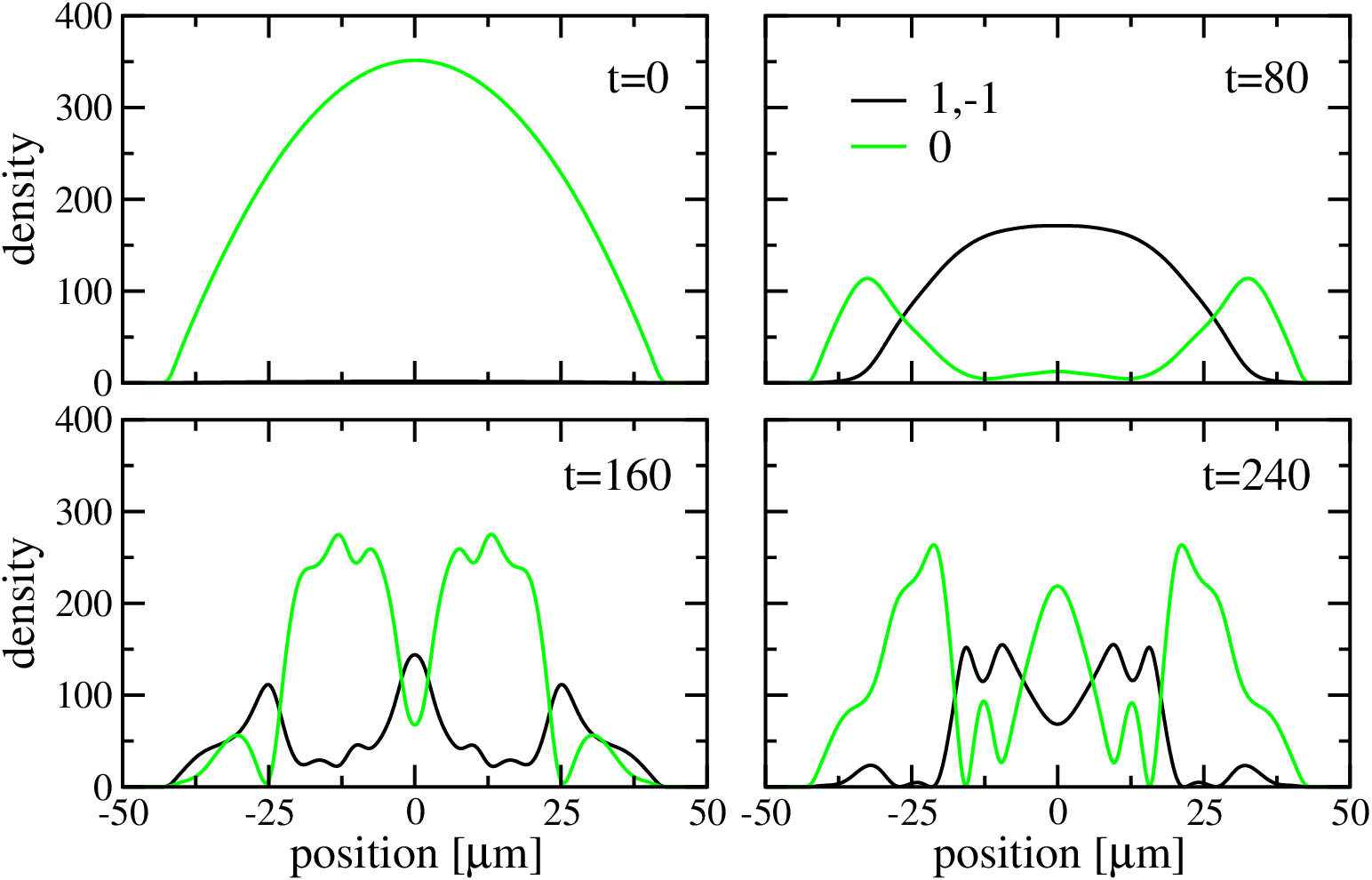}
\caption[]{(Color online) Density profiles of the spin components at different
times (in ms) at $T=0$, $n_1(z,t)$ (black) and $n_0(z,t)$ (green).
The initial configuration corresponds to
$(0.5 \%,99 \%,0.5 \%)$, and $\theta=0$
(dashed line in Fig.\ref{fig1}).
}
\label{fig2}
\end{figure}

In Fig.\ref{fig2} we plot the density
profiles of the spin components for the configuration  $\theta=0$
(see Fig.\ref{fig1}). At the initial stages of the evolution,
the population of the spin-0 component decreases due to the spin
exchange interaction, and thus the $\pm 1$ spin components 
start to be populated by the same amount, keeping the
symmetry of the initial state. The total magnetization is conserved
along the time evolution.
Initially ($t < 100$ ms),
the conversion of atoms from 0 to $\pm$ 1 states mainly occurs
at the central part of the condensate, where the density is higher
and thus the coupling between different spin components is more
effective, see Eq. (\ref{veff}). 

Then, the $\pm$ 1 spin components swing back to the $0$ component, and 
vice versa. The oscillations between the populations of the $m=0$ 
and $m=\pm 1$ states are not regular and present a dynamical
instability around $t \sim 100$ ms, when the large amplitude 
oscillations become small amplitude oscillations 
\cite{Robins01,Saito05,You2005}.
At this moment the condensate starts the multi-domain formation 
process into small dynamical spin domains. A simple estimation of the
time scale for the appearance of the instability $t_{dom}\sim 2 \pi \hbar/(c_2 n)$ 
has been provided 
by studying the normal excitation modes of the system \cite{Saito05,You2005}.
In our case, taking  $n \sim 2 \times 10^{14}$ cm$^{-3}$ 
at the center of the trap, $t_{dom} \sim 140 $ ms, which is in
agreement with our results.
Notice that, in the present study,  the analysis is performed  
directly from  the spatial and temporal evolution  of the population of the 
spin components.

The transfer of population remains coherent along
all the time evolution, and it does not become chaotic even 
for large times.  
The number of small spin domains does not grow indefinitely, but 
it is limited by a characteristic size of the spin domains ($l_{dom}$) that
depends on the internal coupling between different
spin components, and more weakly on temperature.  
The formation of multiple spin domains,
also found recently in Ref.\cite{Saito05,You2005},
is a result of the
ferromagnetic character of the spin-1 $^{87}$Rb condensate:
$c_2<0$ favors the spatial separation of $\pm 1$ atoms from
0-atoms, as obtained in our numerical results.
Since in our simulations the initial condition for $m=\pm 1$
are identical,
the density profiles of $m=\pm 1$ components are equal
along the time evolution. 
Apparently, the value of $l_{dom}\sim 5 ~ \mu$m, being much smaller than
the harmonic oscillator length, 
does not depend 
on the external trapping potential, but it is an intrinsic characteristic 
of the spin coupling strength \cite{You2005}.

\begin{figure}
\includegraphics[width=0.85\columnwidth,
angle=0, clip=true]{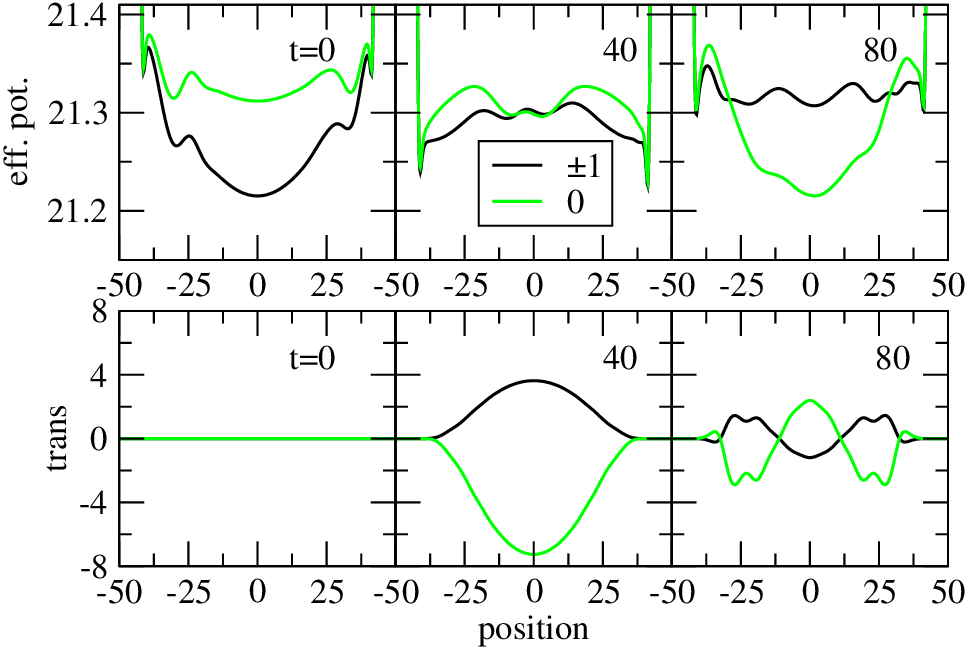}
\caption[]{ (Color online) 
For the initial configuration $(0.5 \%,99 \%,0.5 \%)$,
with $\theta=0$ and $T=0$.
Top panels: Effective potentials of the spin components
for $t=0,40$ and $80$ ms.
Bottom panels: Transfer of populations $\delta \dot{n}_m(z,t)$
for the same times.
}
\label{fig3}
\end{figure}

The dynamical evolution between spin components can be
easily interpreted in terms of the effective potential
$V^{eff}_m(z,t)$ felt by each spin component $\psi_m(z,t)$,
and from the continuity equations.
For the same initial configuration as in Fig.\ref{fig2},
we plot in Fig. \ref{fig3} the effective potentials
$V^{eff}_0$ and $V^{eff}_1=V^{eff}_{-1}$ (top panels),
and the local transfer of population $\delta \dot{n}_0(z,t)$ and
$\delta \dot{n}_1(z,t)=\delta \dot{n}_{-1}(z,t)$ (bottom panels) for
$t=0,40$ and $80$ ms.
For a fixed time $t$, $\delta \dot{n}_m(z,t)$ represents the
local exchange of atoms between spin components. If 
$\delta \dot{n}_m(z,t)< 0$ at that position the 
population of the spin-$m$ component is decreasing and thus
the atoms convert to the other spin components.
For example at the first stages of the time evolution
($t \leq 60$ ms)
$\delta \dot{n}_1(z,t)=\delta \dot{n}_{-1}(z,t)>0$, and consequently
$\delta \dot{n}_0(z,t)<0$: the spin-exchange interaction
is favoring the conversion from 0 to $\pm 1$ states as it 
is shown in Fig.\ref{fig1}. At $t=0$ this is also true, but it 
cannot be appreciated from the scale of the figure.
Moreover, since the minimum
and maximum of $\delta \dot{n}_0(z,t)$ and $\delta \dot{n}_{\pm 1}(z,t)$, 
respectively are at the center of the condensate, the spin-exchange
mainly occurs at the central region, as we have already commented
in Fig.\ref{fig2}.
At $t=80$ ms $\delta \dot{n}_{\pm 1}$ is positive at the boundaries 
of the 
condensate, and negative at the central region, whereas $\delta \dot{n}_0$
has the opposite behavior. Therefore, the population with
$m = \pm 1$ is decreasing at the center and increases at the boundaries, 
and the $\pm 1$ atoms convert to $0$ state mainly at the center, 
where the $m=0$-condensate develops a new central peak.

We have also performed simulations starting from other initial
conditions. In particular, we observe that multi-domain formation
is inhibited if the starting configuration coincides with the
ground state composition ($25 \%, 50\% , 25 \%$). Moreover, we find
a good agreement with the very recent experimental results of
Chang et al. \cite{nature}, starting from (0, 3/4, 1/4) and converging
to (1/5, 2/5, 2/5).


\section{Spinor dynamics at finite $T$}

The spinor dynamics of these multicomponent quantum gases is
rather complex due to the internal coupling between different
spin components. We consider now thermal effects in the
spinor dynamics. 
At low temperature, thermal excitations can be described
within the Bogoliubov-de Gennes  theory. Recently, finite temperature effects 
in the equilibrium  density distribution 
of the condensed and non-condensed components 
of spin-1 trapped atoms has been
investigated within Hartree-Fock-Popov theory \cite{Zhang04}.
We will investigate thermal effects in the spinor dynamics
using a Bogoliubov-de Gennes description of the thermal cloud.
For a highly elongated trap, the condensate
is quasi-1D, and total density fluctuations are 
strongly suppressed at small temperatures, whereas
phase fluctuations are relevant and in the Thomas-Fermi regime can be 
described analytically in terms of Legendre polynomials
$P_j(z)$ \cite{Petrov00,fotka}
Analogously, we assume that spin fluctuations
can be disregarded in a first approximation.

In Refs.\cite{Lewandowski,Tfinita} it has been shown that at finite temperature
each spin component has its own thermal cloud. Thus,
we assume that initially the condensate corresponding to each 
spin-$m$ component can be described by an order parameter
$\psi_m(z)=\sqrt{n_m(z)}\exp(i\phi_m)$, which
has a random fluctuating phase \cite{another}
\begin{eqnarray}
\phi_m(z)&=&[4 n^0(z)]^{-1/2} \sum_{j=1}^\infty
                 \left(\frac{j+1/2}{R_{\rm TF}}\right)^{1/2}
                 \left[\frac{2 \mu}{\epsilon_j} 
        \left(1\right.\right.\nonumber \\
&-&\left.\left. (z/R_{\rm TF})^2 \right) \right]^{1/2}
        P_j(z/R_{\rm TF})\,(a_j^m+a_j^{m*}) \,
\label{Tphase}
\end{eqnarray}
where $n^0(z)$ is the equilibrium total density profile of the
initial condensate, $\mu$ and
$R_{\rm TF}$ are its chemical potential and Thomas-Fermi radius,
$\epsilon_j=\hbar \omega_z \sqrt{j(j+1)/2}$ is the spectrum
of low lying axial excitations \cite{spectrum}, 
and $a_j^m$ ($a_j^{m*}$) are complex amplitudes that replace the
quasi-particle annihilation (creation) operators in the mean-field
approach. In the numerical calculation in order to
reproduce the quantum statistical properties of the phase
\cite{Dettmer01},
$a_j^m$ and $a_j^{m*}$ are sampled as random 
variables with a zero mean value and 
$\langle |a_j^m|^2 \rangle=N_j^m$, where 
$N_j^m=[{\rm exp}(\epsilon_j/k_BT)-1]^{-1}$ is the occupation number
for the quasi-particle mode $j$ \cite{Dettmer02,Kreutzmann03}. 

\begin{figure}
\includegraphics[width=0.85\columnwidth,
angle=0, clip=true]{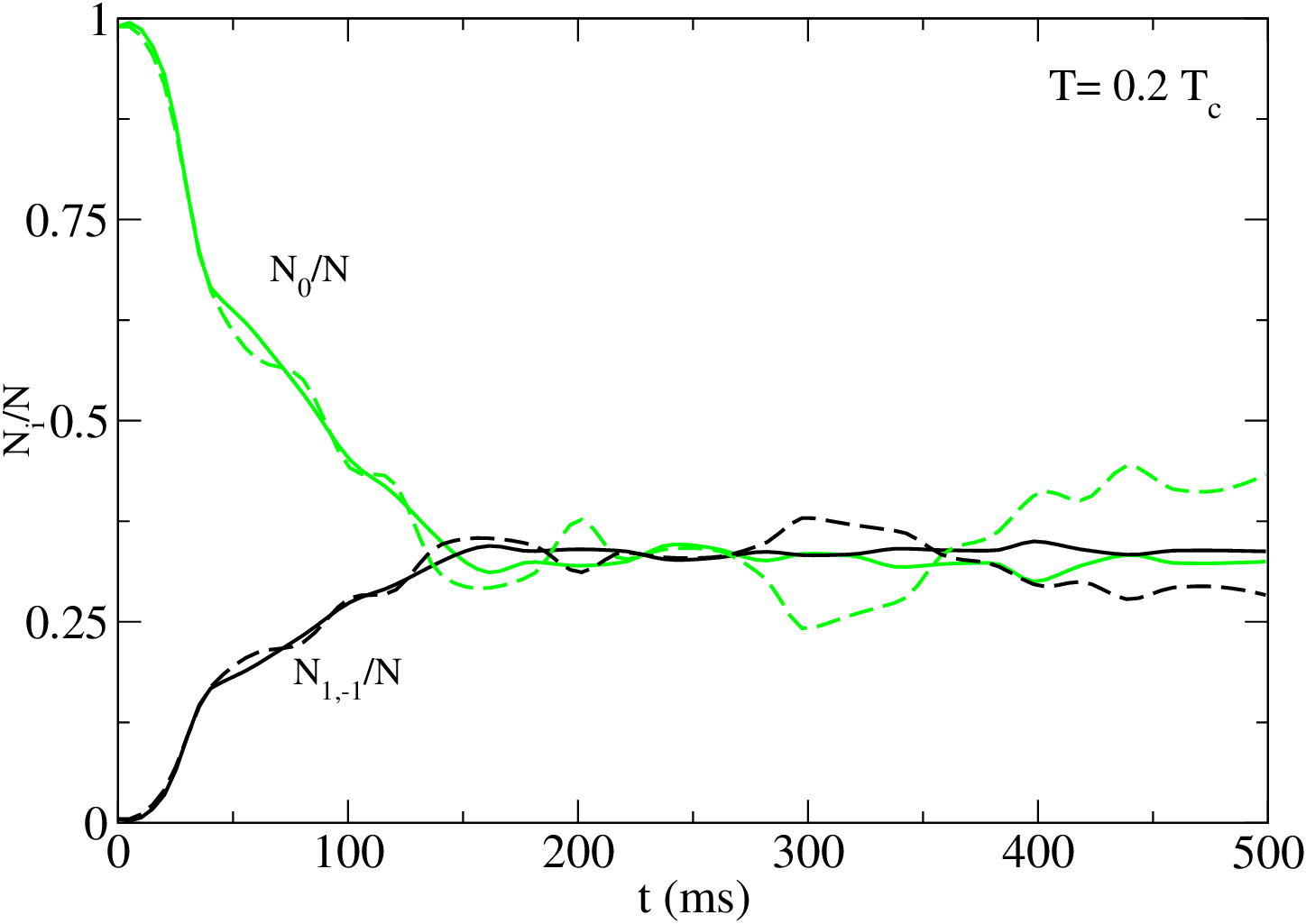}
\caption[]{(Color online) 
Population of the spin components as a function
of time for the initial configuration $(0.5 \%,99 \%,0.5 \%)$
at $T=0.2 T_c$.
Solid lines: numerical results averaged over 20 random
initial relative phases $\theta$. Dashed lines: initial phase
$\theta=0$.
}
\label{fig4}
\end{figure}

In Fig.\ref{fig4} we plot the dynamical evolution of the spin 
populations at $T=0.2 T_c$ for the same initial populations as in Fig.\ref{fig1}. 
The critical temperature of Bose-Einstein condensation, 
$T_c=\hbar \omega_z N/\ln(2N)$,  corresponds here to  a single 
component Bose gas in a harmonic trap of frequency $\omega_z$.  
Solid lines correspond to the numerical
results averaged over 20 random initial relative phases, and 
dashed lines to a single run with initial phase $\theta=0$.
The interaction of the condensate atoms with the thermal clouds
smears out the oscillations present at $T=0$ (Fig.\ref{fig1})
and leads to an asymptotic
configuration with all $m$ components equally populated 
(equipartition). 

\begin{figure}
\includegraphics[width=0.85\columnwidth,
angle=0, clip=true]{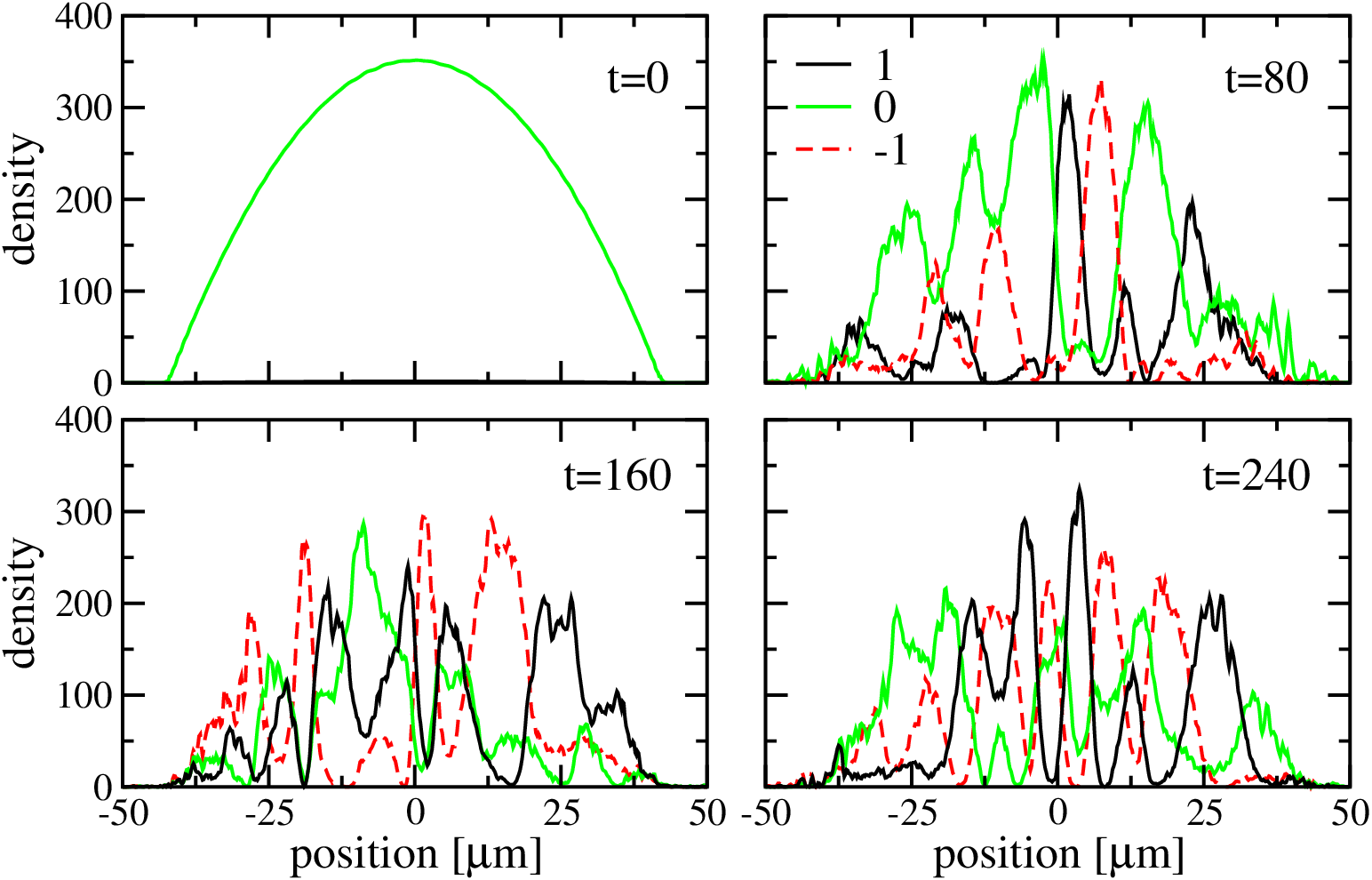}
\caption[]{(Color online) 
Density profiles of the spin components at different
times (in ms) at $T=0.2 T_c$ for the same initial configuration
as in Fig.\ref{fig2}.
}
\label{fig5}
\end{figure}

Thermal effects in the density profiles of the spin components
are shown in Fig.\ref{fig5}.
The occurrence of
phase fluctuations due to thermal excitations at $t=0$
transforms to modulations of the density during the time 
evolution as in a spin-polarized single component elongated
condensate \cite{Dettmer01}. In a spinor condensate this leads
to multi-domain formation at much
earlier times than at $T=0$ (see Fig.\ref{fig2}), 
and the density profiles are no longer symmetric at finite $T$.
Moreover, the existence of different thermal clouds for
each spin component \cite{Tfinita} breaks also
the symmetry of the $m= \pm 1$ spin components of
the initial state of system. Therefore the $m =\pm 1$
density profiles are different during the time evolution,
and the three components
separate in different spin domains. The local
magnetization is not longer conserved as at $T=0$ but,
as expected, the
total magnetization is still a conserved quantity along 
all the time evolution.
Similar results are obtained at lower temperatures $T/T_c=0.01$
and 0.001.

\section{Conclusions}

In this paper,
we have studied the spinor dynamics of a spin-1 $^{87}$Rb 
condensate in a highly elongated trap. We have solved the
full three coupled dynamical equations for the spin components
within Gross-Pitaevskii framework without any further 
approximations. This is in fact necessary, since approximated approaches frequently 
mask some of the aspects of the dynamics.
We have also considered here finite temperature 
effects, using the approach of Ref. \cite{Dettmer02}. 
We have found that
the spinor dynamics towards the steady state is not monotonic
but rather slowly damped, involving a coherent transfer of population
between different spin components. At finite temperature the
coherent oscillations of populations  are almost smeared out. The internal coupling
of the spin components leads to the  formation 
of multiple  spin domains of a small, but finite 
characteristic length $l_{dom}$, which does not decrease with time, and seems thus to be determined 
intrinsically by the nonlinear interactions.
This scale is evidently larger than the 
condensate healing length, 
$l_{heal}=2 \pi \hbar/\sqrt{2 M c_0 n}$, 
which for scattering lengths $a$ of order of 5 nm 
is of order of 10-100 nm. 
In fact, $l_{dom}$ is of the order of the spin healing length
$\xi_{s}=2 \pi \hbar/\sqrt{2 M |c_2| n}$.
 The presence of different thermal clouds
for each spin component breaks the symmetry of the $m=\pm 1$ components,
and therefore separates them in different spin domains.
For a condensate with initially zero magnetization,
the spin populations oscillate around the ground state
configuration $(25 \%,50 \%,25 \%)$ at $T=0$, whereas at finite 
temperature the interaction of the condensate atoms with their corresponding
thermal clouds leads to equipartition in populations, i.e. 
$(1/3,1/3,1/3)$.

Our results shed, in our opinion, also some light on the question of decoherence. Our simulations, 
 and in particular the finding of multi-domain formation suggest that 
decoherence undergoes enhancement with the number of
components in the system. 
Of course, there are many open questions connected to this, e.g. does the
multi-domain formation  go along with a loss of phase relations,  and give rise to some
enhanced (generalized) phase fluctuations?  
These questions go beyond the present study, 
and will be discussed in detail 
in a future publication, in which we will compare the $F=1$ and $F=2$ case.

This research was partially supported by DGICYT (Spain) Project No. 
BFM2002-01868, MEC (Spain) Project FIS2005-04627, Generalitat de Catalunya Project No. 2001SGR00064,
Deutsche Forschungsgemeinschaft (SFB 407, SPP 1116),  
and the European Science Foundation programme QUDEDIS. J. Mur-Petit acknowledges also 
support from the Generalitat de Catalunya. 
 M. G. thanks the ``Ram\'on y Cajal'' Program (Spain) for
financial support.
We are grateful to Prof. T. Pfau, Prof. H. T. C. Stoof and Dr. I. Carusotto  
for discussions.

\newpage


\begin{thebibliography}{99}
\bibitem{fotka} Strictly speaking, in quasi-1D condensates, the phase fluctuations at 
$T=0$ are
described by Jacobi polynomials (for details see \cite{Petrov00,spectrum}), in contrast to 
exactly 1D condensates. Already at  $T\simeq 0.2 T_c$, however, quantitative difference between thermal effects 
in quasi-1D and strictly 1D cases becomes irrelevant (compare \cite{Dettmer02,Kreutzmann03}).
\bibitem{another} In general, Bogoliubov-de Gennes equations describing the 
thermal quasiparticle 
excitations are derived as linearizations of the coupled Gross-Pitaevski equations, and as such
they   mix various spin components. If, however, the linearization is performed around a practically 
fully polarized state  (such as the ones considered in this paper), the fluctuations of the large  
spin component decouple from the other components. The latter decouple as well, since  they 
are initially very 
weakly populated, and their dynamics is dominated by the mean field spin independent effects 
caused by the large spin component. This is particularly true for $|c_2|\ll c_0$.
\end{thebibliography}
\end{document}